\documentclass[aps,pre,twocolumn,showpacs,superscriptaddress]{revtex4-1}
\usepackage{amsmath,amssymb}
\usepackage{graphics,graphicx,color}
\usepackage{dcolumn,bm}
\usepackage{tikz}
\usepackage{natbib} 
\usepackage[utf8]{inputenc}
\usepackage[colorlinks=true,citecolor=blue,urlcolor=blue]{hyperref}

\def\[{\left[}
\def\]{\right]}
\def\({\left(}
\def\){\right)}
\def\be{\begin{equation}}
\def\ee{\end{equation}}
\def\bea{\begin{eqnarray}}
\def\eea{\end{eqnarray}}

\newcommand{\gaug}
{\affiliation{Institute for Theoretical Physics, Georg-August-Universit\"at G\"ottingen, 37077 G\"ottingen, Germany}}

\begin{document}
\title{Interaction from Structure using Machine Learning: in and out of Equilibrium}

\author{Saientan Bag}
\email[Email: ]{saientan.bag@kit.edu }
\affiliation{Institute of Nanotechnology, Karlsruhe Institute of Technology, Karlsruhe, Germany}

\author{Rituparno Mandal}%
\email[Email: ]{rituparno.mandal@uni-goettingen.de}
\gaug

\begin{abstract}

Prediction of pair potential given a typical configuration of an interacting classical system is a difficult inverse problem. There exists no exact result that can predict the potential given the structural information. We demonstrate that using machine learning (ML) one can get a quick but accurate answer to the question:``which pair potential lead to the given structure (represented by pair correlation function)?" We use artificial neural network (NN) to address this question and show that this ML technique is capable of providing very accurate prediction of pair potential irrespective of whether the system is in a crystalline, liquid or gas phase. We show that the trained network works well for sample system configurations taken from both equilibrium and out of equilibrium simulations (active matter systems) when the later is mapped to an effective equilibrium system with a modified potential. We show that the ML prediction about the effective interaction for the active system is not only useful to make prediction about the MIPS (motility induced phase separation) phase but also identifies the transition towards this state.

\end{abstract}


\maketitle

\maketitle

\section{Introduction}

One of the basic questions in statistical mechanics is what structure a system of interacting particles will attain given a microscopic pair wise interaction at a given temperature. Typically this kind of questions are addressed either by Molecular Dynamics (MD) or Monte Carlo simulation or some semi-analytical approach like integral equation theory (see ~\cite{evans79,howells72,johnson63,johnson64,banerjee74,hansen90} for a non exhaustive list of references and ~\cite{toth07} for a detailed review) which enables one to calculate the pair correlation function $g(r)$ or structure factor $S(q)$ (related to $g(r)$ through Fourier Transform) when the pairwise interaction potential $V(r)$ is known. On the other hand, an inverse problem is the prediction of pair potential from a typical configuration of system of particles (or a structural correlation like $S(q)$ or $g(r)$). The inverse theorem\cite{henderson1974uniqueness} says that for the fluids with only pairwise interaction (quantum or classical), the pair potential $V(r)$ that leads to a specific $g(r)$ is unique and  it makes the above mentioned question a well defined one. 
\begin{figure}
\begin{center}
\includegraphics[width=.85 \columnwidth]{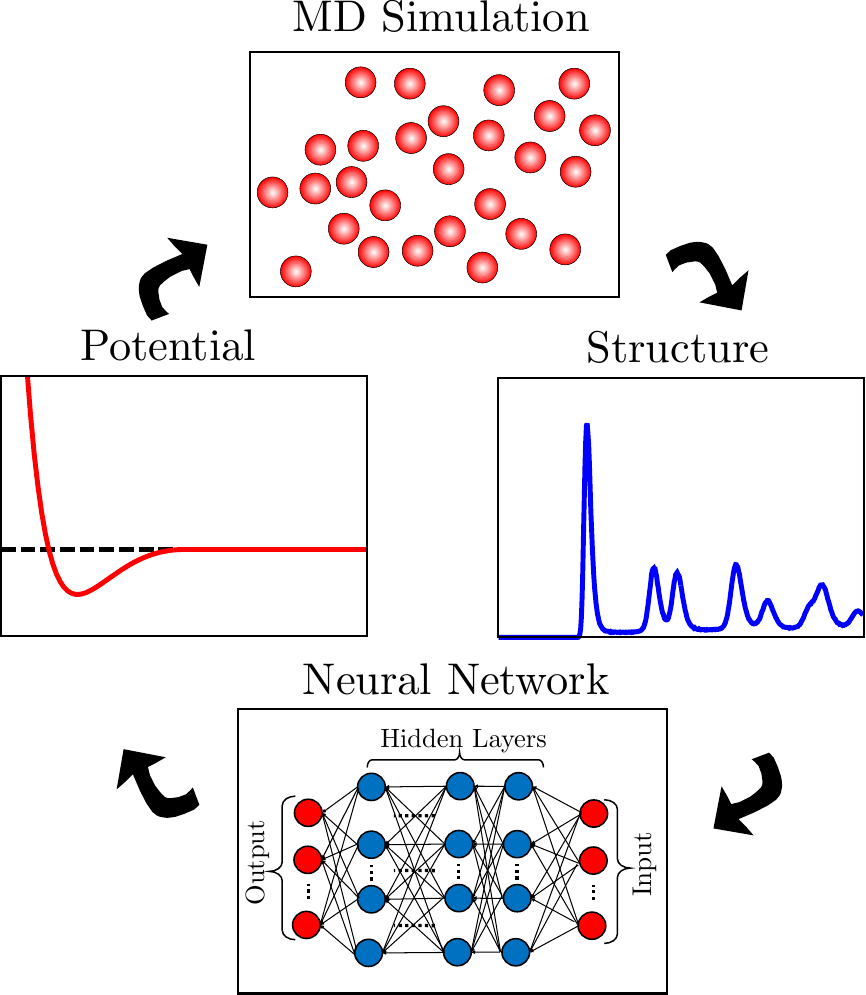}
\caption{(Color online) The usual MD simulation gives structure ({\textit{e.g.}} pair correlation function $g(r)$) as an output given a pair potential $V(r)$ at a fixed temperature and density. The above shows the conventional route to go to structure given the potential through MD and below shows the inverse problem of asking which pair potential lead to the observed pair correlation function using neural network (NN).}
\label{fig:schematic}
\end{center}
\end{figure}
 This theorem has been proved multiple times in different context\cite{henderson1974uniqueness,baranyai2003statisztikus, zwicker1990does,tothbara}. By construction the theorem proves the uniqueness but does not provide a prescription to connect structure to interaction. Initial exploration in this direction was started by Johnson, Hutchinson and March\cite{johnson63,johnson64} where they used Born-Green (BG) hierarchy and Percus Yevick theory to get an approximate result. Several other methods also have been proposed along a similar line. To name a few the iterative method by Schommer and Soper\cite{schommers1983pair,schommers1973pair} , the LWR scheme by Levesque, Weis and Reatto\cite{levesque1985pair}, Reverse Monte Carlo method\cite{toth1999comparison,toth2000direct} and Force Matching Method\cite{izvekov2004effective, ercolessi1994interatomic}  etc. were discovered over the years. But till today, there exits no exact result that can answer this question. Instead most of the methods described before, do it in a approximate manner using iterative refinements of the pair potential to achieve a desired structure. These existing methods are numerically quite involved and in many cases being iterative computationally complex. Using a completely different route, we addressed the above mentioned {\textit{inverse problem}} using machine learning tools.

In the past few years, ML approaches have been extensively used to address novel scientific problems in almost all branches of research~\cite{carleo2019machine}. To find the mass of the galaxy cluster~\cite{ntampaka2015machine,ntampaka2016dynamical} in astrophysics/cosmology, generate  an accurate force field(FF) in chemical physics~\cite{sifain2018discovering} or  to learn the new phases of a  certain quantum system~\cite{carrasquilla2017machine} the  ML methods have been  proved to be of unprecedented use. Use of ML tools in this problem makes it addressable without going into involved mathematical or numerical schemes. The benefit of the ML approach is also that it is simple and accurate and can be used easily as a numerical tool bypassing complex iterative numerical methods. Such neural network based "quick and dirty" method has been explored before~\cite{toth05} on liquid phase, to predict the pair interaction from the structure factor $S(q)$ of a liquid. Instead we used pair correlation function $g(r)$ as training feature for our NN and have showed that this method is also extendable for crystal as well as gaseous phase.

We further show the validity and effectiveness of the method in out of equilibrium scenario (namely in an active matter system). To establish that we first show the effectiveness and accuracy of the ML method for a family of potentials where the degree of attractiveness can be tuned quite easily. Using MD simulation we generated pair correlation functions from equilibrium snapshots from each such pair potential at a particular density and temperature. We first make sure this process of learning works using neural network and it can predict with desired accuracy. Once the success of the method has been demonstrated for the equilibrium systems with different phases (crystal, liquid and gas) we extend the same methodology for a specific out of equilibrium system known as active matter~\cite{sriram10,marchetti13,elgeti15,lowen16}. We take a canonical and well known example of such active matter system: active brownian particle (ABP) model. This model system ~\cite{marchetti12,takatori15,levis17,solon18,tailleur15} has been used quite extensively to model the motion of self propelled Janus colloid, bacteria etc. and the model shows a novel phase separation process known as motility induced phase separation (MIPS).  Estimation of effective two body interaction (when mapped into an equilibrium problem) for ABP has been worked out in many recent works~\cite{brader15,brader18}. Mapping an active matter system in general~\cite{brader15,fodor16,maggi16,evans16,tailleur2020} or a living system~\cite{bordeu20} to an effective equilibrium scenario to understand the underlying emergent interaction has generated significant interest. Hence using an alternative and generic path, we have shown that the ML based method works well for out of equilibrium systems; especially it can predict quite correctly the transition from gaseous to MIPS like phase and the effective interaction required to observe MIPS. We also verified the accuracy of the prediction by comparing the pair correlation function of the active particle (ABP) simulation and the passive simulation with predicted effective pair potential. 

\begin{figure}
\begin{center}
\includegraphics[width=.9 \columnwidth]{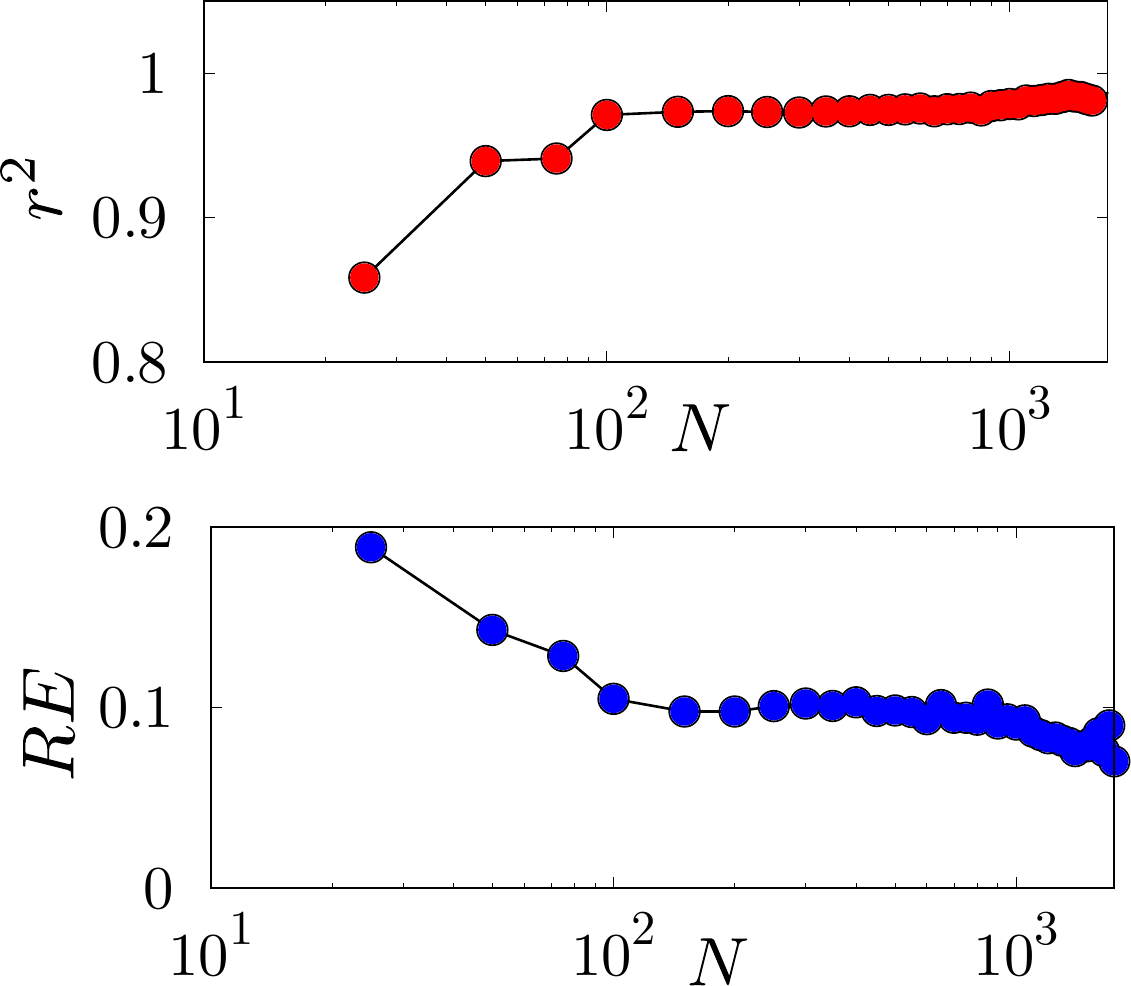}
\caption{(Color online) Learning curve for the NN based ML model studied: (top) Accuracy ($r^2$) and (bottom) relative error ($RE$) of the test data set (see text for details), as a function of number of data ($N$) in the training set. Both $r^2$ and $RE$ saturate to a value of $\sim 0.975$ and $\sim 0.09$ respectively indicating a fairly accurate ML model within few hundred data points in the training set.}
\label{fig:accuracy}
\end{center}
\end{figure}

The rest of the paper is organized as follows. In Sec. II we introduce basic methodology for training data generation using MD and about the scheme of machine learning. In Sec. III we describe the results with three subsections:A, B, C. In subsection A we discuss the training and testing of the ML model. In subsection B we discuss the results for the equilibrium system . In subsection C we analyse the results from the out of equilibrium (ABP) system which includes qualitative prediction about the crossover to the MIPS phase. In this subsection, we also discuss the predicted effective potential for the active system along with the quality of the predictions. Finally, in Sec. IV we conclude with a summary and a discussion. Some technical details of the machine learning algorithms are described in the the Appendixes.
\begin{figure*}
\begin{center}
\includegraphics[width=.95 \textwidth]{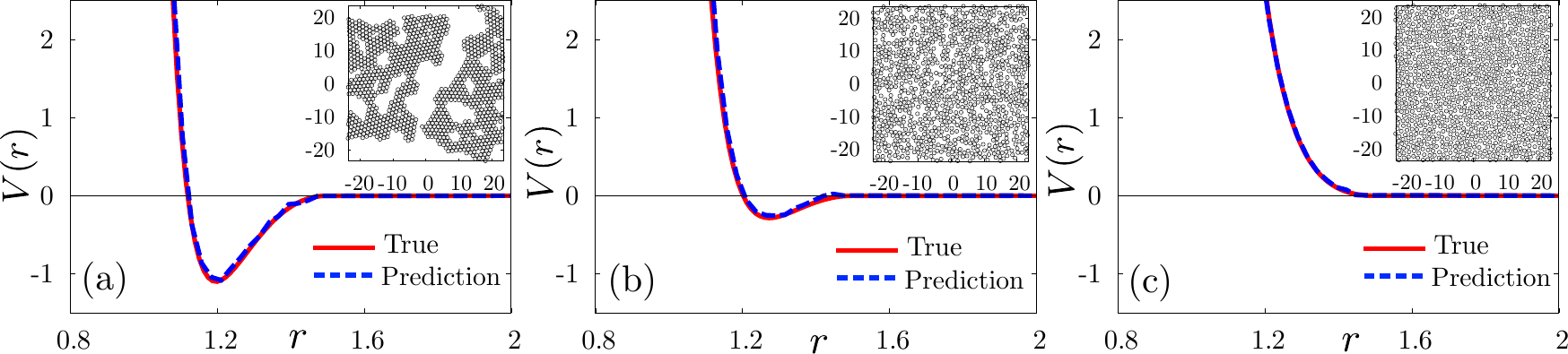}
\caption{(Color online) Comparison between the true potential (red solid line) and the predicted potential using the ML algorithm for (a) crystalline state (b) liquid like state (c) gaseous state, where attractive part in the potential is relatively strong, moderate and identically equal to zero correspondingly. Inset shows typical configurations (for this potential) from the simulation for each cases, {\textit{i.e.}} insets in (a), (b) and (c) shows crystalline state, liquid like state and gaseous state respectively.}
\label{fig:passive_compare}
\end{center}
\end{figure*}

\section{Methodology}

As detailed in Fig.~\ref{fig:schematic}, we first generate the training data for the ML model by performing a set of equilibrium molecular dynamics simulation with different interaction potentials $V(r)$ at a fixed temperature and density. We then calculate the average pair correlation function for each parameter set (representing the corresponding potential) from those equilibrium configurations.This above formalism of getting structural information from potential (top part of the panel in Fig.~\ref{fig:schematic}) is one of the standard problems in statistical mechanics and condensed matter physics. Now to implement the reverse route (bottom part of the panel in Fig.~\ref{fig:schematic}) we use pair correlation function as training data for a supervised machine learning scheme. The supervised ML was implemented using a neural network (NN) model. The details of the steps are described below.
\subsection{MD Simulation}\label{MD Simulation}
The molecular dynamics simulations have been carried out in a two dimensional domain with $N_p=1000-4000$ particles. We use usual periodic boundary conditions (PBC). We implement Brownian dynamics scheme using simple Euler integrator with time step $\Delta t=2 \times 10^{-4}$. The Equation of motion reads as 
\begin{equation}
     \frac{\mathrm{d}\mathbf{r}_i}{\mathrm{d}t}
    = - \frac{1}{\zeta}\sum \limits_{j \ne i}
     \frac{\partial V \left (\left |\mathbf{r}_i-\mathbf{r}_j \right | \right )}
          {\partial \mathbf{r}_i} + {\boldsymbol{\eta}}_i(t)
 \label{eq:dridt}
\end{equation}
where $\mathbf{r}_i$ is the position vector of the $i$ th particle, $\zeta$ is the friction coefficient, $V_{I}$  is the interaction potential. The thermal noise,
${\boldsymbol{\eta}}_i(t)$, has zero average and is delta-correlated {\textit{i.e.}} $\langle {\boldsymbol{\eta}}_i(t) {\boldsymbol{\eta}}_j(t^{\prime}) \rangle= 2 D \delta_{ij} \delta(t-t^{\prime}) \mathbf{I}$ 
where $D$ is the translational diffusion
coefficient $D=\frac{k_B T}{\zeta}$ and $\mathbf{I}$ is the identity matrix. We keep the temperature fixed at $T=0.2$, $\zeta=1.0$ and maintained the area fraction $\phi=0.45$.
 For the purpose of training we parametrise the potential using the form:
 \begin{equation}
    V
    =4 \epsilon \left[ \left(\frac{\sigma}{r_{ij}} \right)^a-\lambda \left(\frac{\sigma}{r_{ij}} \right)^{b}\right]
 \label{eq:pot}
\end{equation}
where $\epsilon$ and $\sigma$ set the scale for energy and length respectively. We have set the value of $\epsilon=10$, $\sigma=1.0$ for all the training simulation result described here. All the potentials has been truncated and shifted at $r_c=1.5\sigma$ such that both the potential and force remain continuous at $r=r_c$ and this has been achieved by a second order smoothening function. Note that by choosing $a$, $b$ (which controls the stiffness of the repulsive and attractive part of the potential respectively) and $\lambda$ which controls the relative strength of the attractive versus repulsive interaction) carefully, we can generate a family of pair potentials which include completely repulsive, moderately attractive to strongly attractive pair interactions. During the MD we allow the system to reach the steady state (by allowing $\sim 10^8$ MD steps) and then collect $100$ equally spaced equilibrium configurations from each trajectory (over a period of $\sim 10^8$ MD steps). All the structural correlations $g(r)$ are time average over $100$ such equilibrium configurations.

\subsection{Machine Learning}
The structure of the resulting MD configurations were 
quantified by calculating pair correlation function ($g(r)$) defined as follows,
\begin{equation}
    g(r)
    =\frac{A}{2\pi r N^{2}} \langle \sum_{i} \sum_{j(i\neq j)} \delta (r-r_{ij})\rangle
 \label{eq:gr}
\end{equation}
Here, $r_{ij}=|\mathbf{r}_{i}-\mathbf{r}_{j}|$, where  $\mathbf{r}_{i}$ and $\mathbf{r}_{j}$ are the position vectors of the
$i$th and $j$th particle respectively. 
$A$ is the area of the simulation box and $N$ is the total 
number of particle simulated. The angular bracket $\langle ... \rangle$ represents time average over independent equilibrium snapshots, as mentioned before.
\begin{figure*}
\begin{center}
\includegraphics[width=.95 \textwidth]{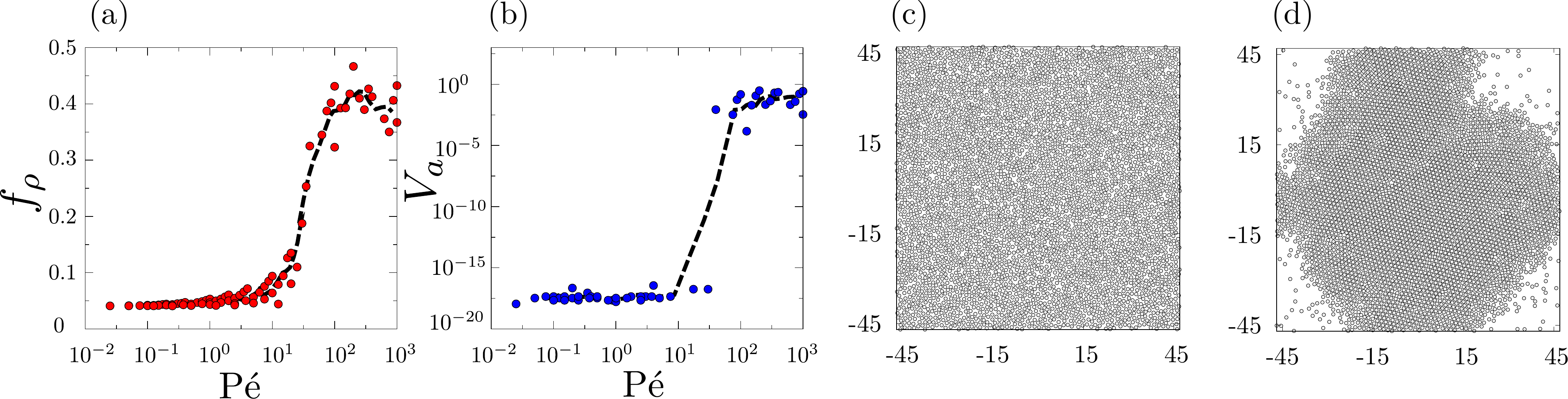}
\caption{(Color online) (a) Segregation order parameter $f_{\rho}$, that characterises mesoscopic fluctuations in the system shows crossover from a homogeneous steady state (see (c) for the snapshot) to a motility induced phase separated state (see (d) for the snapshot). (b) Strength of the attractive interaction or $V_a$ (measured from the predicted potentials from ML) also show a similar crossover at $\text{P\'e} \sim 100$.}
\label{fig:active_compare}
\end{center}
\end{figure*}
We further train a NN to predict the underlying potential in which MD simulation was performed given the pair correlation function as input (see Figure \ref{fig:schematic}).  
We train a NN having $3$ hidden layers with $1000$ nodes in each layer. “Leaky ReLu” activation function was used for all $3$ fully connected hidden layers. “Linear” activation function was used for the output layer.  “L1” regularization scheme was further used to prevent over fitting. The pair correlation function was binned to generated $1000$ points while the potential was discretized in $n_0=20$ points. Therefore, our machine learning model had $1000$ dimensional input and 20 dimensional output. We have used the open source library KERAS\cite{chollet2015keras} for the implementation of the NN. We have also explored the case with larger output vector (with $n_0=40$ points) for this and obtained similar results (for details see Appendix A and Fig.~\ref{fig:s1}).

\section{Results}

\subsection{Training and Testing of the Machine Learning Model}
As previously mentioned, we first do passive simulations for $\sim2000$ independent pair potentials, parametrised by few constants (see ~\ref{MD Simulation} for details). We chose pair correlation function $g(r)$ to be the training feature. The entire data set for the NN training was divided into training and test data set. To generate the learning curve (see Fig.~\ref{fig:accuracy}) of our model, we gradually increase the number of data points in the training data set and evaluate the quality  of the trained model on a data set with 100 points (test data). The quality of the trained model was assessed by calculating the accuracy ($r^{2}$)(see Eq.~\ref{eq:r2}) and relative error ($RE$) (see Eq.~\ref{eq:re}) of the NN  prediction over the actual value (see Fig.~\ref{fig:accuracy}). For that we first calculate mean absolute error and accuracy ($r^2(j)$ and $MAE(j)$) respectively for $j$-th component of the output vector with $j \in [1, n_0]$ as
\begin{equation}
    r^{2}(j)
    =1-\frac{\sum_{i=1}^{M}{(y^j_{i}-f^j_{i})^{2}}}{\sum_{i=1}^{M}{(y^j_{i}-\langle y_{i} \rangle^j
)^{2}}}
\end{equation}
and
\begin{equation}
    MAE(j) 
    =\frac{\sum_{i=1}^{M}{|(y^j_{i}-f^j_{i})|}}{M}
\end{equation}
where $y^j_{i}$ is the  NN predicted value and $f^j_{i}$ is  the actual value for the $j$-th component of the output vector for $i$-th test data and $\langle y_i \rangle^j= \frac{1}{M}\sum_{i=1}^{M}y_{i}^{j}$. Also, $M$ is the total number of data points in the test set and  $n_0$ is the dimension of the output vector. Finally we define $r^2$ and $RE$ as, 
\begin{equation}
r^2=\frac{1}{n_0}\sum_{j=1}^{n_0} r^2(j)
 \label{eq:r2}
\end{equation}
\begin{equation}
RE=(n_0 \bar{f})^{-1}\sum_{j=1}^{n_0} MAE(j)
 \label{eq:re}
\end{equation}
where $\bar{f}=\frac{1}{Mn_0} \sum_{i=1}^{M} \sum_{j=1}^{n_0} |f_{i}^{j}|$.
As evident from the learning curve (see Fig.~\ref{fig:accuracy}), a fairly accurate model is achieved within few hundred data points in the training set (see Appendix C and Fig.~\ref{fig:s3} and for Fig.~\ref{fig:s4} the details about optimal fitting). We have also explored {\textit{Random Forest (RF)}} machine learning model (see Appendix B and Fig.~\ref{fig:s2} for details) but this yields relatively poor results in terms of accuracy and error.

\subsection{Equilibrium System}

To exhibit  the predictive power of the trained NN, we plot the predicted potential, together with the actual ones for three different test cases (gas, liquid and crystal phase). In Fig.~\ref{fig:passive_compare}, we demonstrate the accuracy of the prediction by comparing the true potential (red solid line) and the predicted potential (blue dashed line) obtained using the NN. As shown in the Fig.~\ref{fig:passive_compare}, our ML model works quite well irrespective of the phase of the system {\textit{i.e.}} liquid, gas and crystal (for the snapshots of the system in  corresponding phases see insets of Fig.~\ref{fig:passive_compare}).

For simplicity our methodology is explained here in two dimensions as a paradigmatic case. But, in two dimensions true crystalline order is impossible to attain at any finite temperature because the low energy excitations which will kill any long range continuous  translational order~\cite{mermin66}. However, it is relatively straightforward to extend our methodology for three or higher dimensional systems where such pathology is absent.

\subsection{Non-Equilibrium System}

As a canonical example of non-equilibrium system we use the active brownian particle (ABP) system. This model has been extensively studied in the active matter literature~\cite{marchetti12,takatori15,levis17,solon18}  especially to understand the motility induced phase separation (MIPS). In this article we would like to understand whether it is possible to estimate the effective potential that would lead to structures similar to the one sampled from the active system. 
The equation of motion for the active particle has an additional term,
\begin{equation}
     \frac{\mathrm{d}\mathbf{r}_i}{\mathrm{d}t}
    = - \frac{1}{\zeta}\sum \limits_{j \ne i}
     \frac{\partial V_{I} \left (\left |\mathbf{r}_i-\mathbf{r}_j \right | \right )}
          {\partial \mathbf{r}_i} + f \hat{\mathbf{n}}_i+ {\boldsymbol{\eta}}_i(t)
 \label{eq:dridt}
\end{equation}
where $\hat{\mathbf{n}}_i$ is the unit vector associated with the orientation of active forcing or propulsion associated with the $i$ th particle. The propulsion force has magnitude $f$ and $\hat{\mathbf{n}}_i \equiv \{\cos{\theta_i},\sin{\theta_i}\}$ where $\theta_i$ is the angle along which the active forcing is acting on the $i$-th particle. The dynamics is diffusive for $\theta_i$ with $D_{\theta}=\frac{1}{\tau_p}$. The net activity in such a system can be measured by a dimensionless quantity: P\'eclet number where $\text{P\'e}=\frac{f \tau_p}{\zeta \sigma}$. For small $\text{P\'e}$ one can see uniform or homogeneous phase and for large  $\text{P\'e}$ the system phase segregates. As the underlying potential for the active simulation is strictly repulsive the clustering can be imagined to be appearing from an effective attraction~\cite{brader15,speck16,tailleur2020} generated from the persistent active forces. We want to estimate this effective potential and then would like to test the results through the passive simulation where we incorporate the predicted potential. 

We first analysed whether the ML algorithm qualitatively works or not. For that we performed non-equilibrium simulations for different value of $f$ and $\tau_p$ such that we can cover a large range ($10^{-2} \le \text{P\'e} \le 1000 $) in P\'eclet number to see the crossover from a gas of active particles to the motility induced phase separated state. The order parameter which measures the degree of segregation or spatial density inhomogeneity is defined as,
\begin{equation}
    f_{\rho}=\frac{\sqrt{\langle n^2 \rangle-{\langle n \rangle}^2}}{{\langle n \rangle}}
 \label{eq:orderp}
\end{equation}
where $n$ is the number of particles in a finite size area element inside the simulation domain. For this computation we have divided our system into $100$ boxes (where each box represents one area element) giving rise to $\langle n \rangle=40$ (for total number of particles $N_p=4000$). The order parameter $f_{\rho}$ clearly captures (see Fig.~\ref{fig:active_compare}) the transition from active homogeneous gas state (with low $f_\rho$) to the MIPS (with high $f_\rho$) as a function of increasing $\text{P\'e}$. To correlate that with the prediction from ML algorithm  we quantified the net strength of the attractive part ($V_a$) (Eq.~\ref{eq:valpha}) of the predicted potential defined as follows (see SI for more details).
\begin{equation}
    V_{a}
    =\int_{0}^{\infty } W(r)V(r)dr 
 \label{eq:valpha}
\end{equation}
where $V(r)$ is the predicted two body potential from the NN and $W(r)=1$ if $V(r)<0$ and $W(r)=0$ otherwise.
If we plot the attractive part ($V_a$) of the predicted potential as a function of P\'eclet number $\text{P\'e}$ we see a crossover from $V_a \sim 0$ to finite $V_a$ (see Fig.~\ref{fig:active_compare}). Note that a non-zero, finite $V_a$ suggests an effective attractive potential therefore a phase separated state (provided the system is below the critical temperature $T_c$ for phase segregation) and a gaseous state otherwise. This similarity (see Fig.~\ref{fig:active_compare} (a) and (b)) in the crossover suggests that the prediction about the effective pair potential is qualitatively giving us the correct picture of the transition to MIPS state.

\begin{figure}
\begin{center}
\includegraphics[width=.94 \columnwidth]{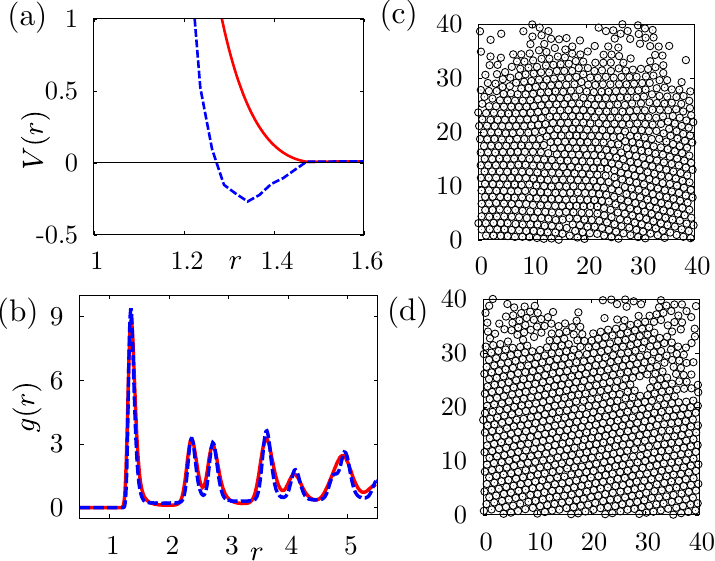}
\caption{(Color online) (a) The bare interaction potential (solid red line) used for ABP simulation and the predicted effective potential (blue dashed line) (b) comparison between pair correlation function $g(r)$ from MIPS (red solid line) and passive simulation with NN predicted potential (blue dashed line). Section of the snapshots taken from MIPS (c) and passive simulation (d) with NN predicted potential for a visual comparison between the states.}
\label{fig:mips}
\end{center}
\end{figure}

We then take one step ahead to ask the question how accurate is the prediction if we focus on a single parameter (say a fixed P\'eclet number, {\textit{i.e.}} the case of a fixed value of $f$ and $\tau_p$). We calculated pair correlation function $g(r)$ from the steady state snapshots generated from the active particle simulation done with only repulsive interaction. We used this $g(r)$ to predict the effective interaction potential using our trained neural network. The predicted pair potential (see Fig.~\ref{fig:mips}) shows clear signature of attractive interaction (blue dashed line represents the NN predicted effective potential). We then do a passive simulation with this predicted potential to check the quality of the prediction. The comparison between the $g(r)$ of the active simulation with the passive simulation with ML predicted pair potential shows (see Fig.~\ref{fig:mips} (b)) the quality of the prediction (for comparison between the configurations from these two simulations see Fig.~\ref{fig:mips} (c), (d)).

\section{Discussion \& Conclusion}

Here in this article we address the problem of predicting pair potential from static structure (by using pair correlation function $g(r)$) using machine learning tools. We show that the multi-layer neural network can be trained to predict the potential quite accurately in equilibrium scenario for all phases: crystal, liquid and gas. We then extend our approach for active matter problems to demonstrate its accuracy and effectiveness in predicting the MIPS transition and effective potential for MIPS like phases. Note that our mapping to equilibrium approach using ML, cannot really understand
the dynamical non-equilibrium properties of active matter. Indeed, there are a plethora of non-equilibrium
phenomena that a passive system with an effective attractive interaction cannot reproduce\cite{vicsek95,lowen15,lowen19,leomach19a,mandal20,lorenzo20}. In future we plan to extend our methods to explore similar dynamical aspects of active matter systems using ML methods (see ~\cite{tociu20} where a similar question has been addressed very recently). We also plan to extend this method to include more complicated cases like binary mixture or poly-disperse systems. It also remains as an open questions, whether the above mentioned approach can be extended for three body or higher order interactions~\cite{turci20}. The parametrizations of the potential represents a family of pair potentials, which we believe covers most of the cases, but yet not complete. But it is straight forward to include other potentials as well using a relatively generic parametrization. To increase the accuracy to even higher degree one can in principle consider a bigger and much diverse data set in terms potential and modelling the active dynamics. Our result will be of interest for structure to pair potential mapping problems in material science and for colloidal systems where complex pair wise effective interaction can be predicted using such black box like approach and also in out of equilibrium problems in active matter or living systems~\cite{brader15, tailleur2020,bordeu20}.

\section{Acknowledgement} We are grateful to Chandan Dasgupta, Debsankar Banerjee, Corneel Casert and Lorenzo Caprini  for insightful discussions and for their valuable comments about the manuscript. This project has received funding from the European Union’s Horizon 2020 research and innovation programme under the Marie Sk\l odowska-Curie grant agreement No 893128.

\appendix 
\setcounter{figure}{0} \renewcommand{\thefigure}{A.\arabic{figure}} 

\section{NN Learning Curve for $n_0=40$}

The neural network based machine learning model presented in the main text of the article had $20$ dimensional output ({\textit{i.e.}} the potential was discretized in $20$ points). One question that might arise naturally is how good the performance will be when we use more number of points for discretization. Here we show that, a neural network based machine learning model with similar accuracy can also be achieved when the output the NN is $40$ dimensional ({\textit{i.e.}} when the potential is discretized in $40$ points). This demonstrates that the results are robust with respect to variation in $n_0$ or the dimensionality of the output vector.The learning curve (both accuracy $r^2$ and relative error $RE$) of the corresponding ML model is shown below.

\begin{figure}
\includegraphics[width=1 \columnwidth]{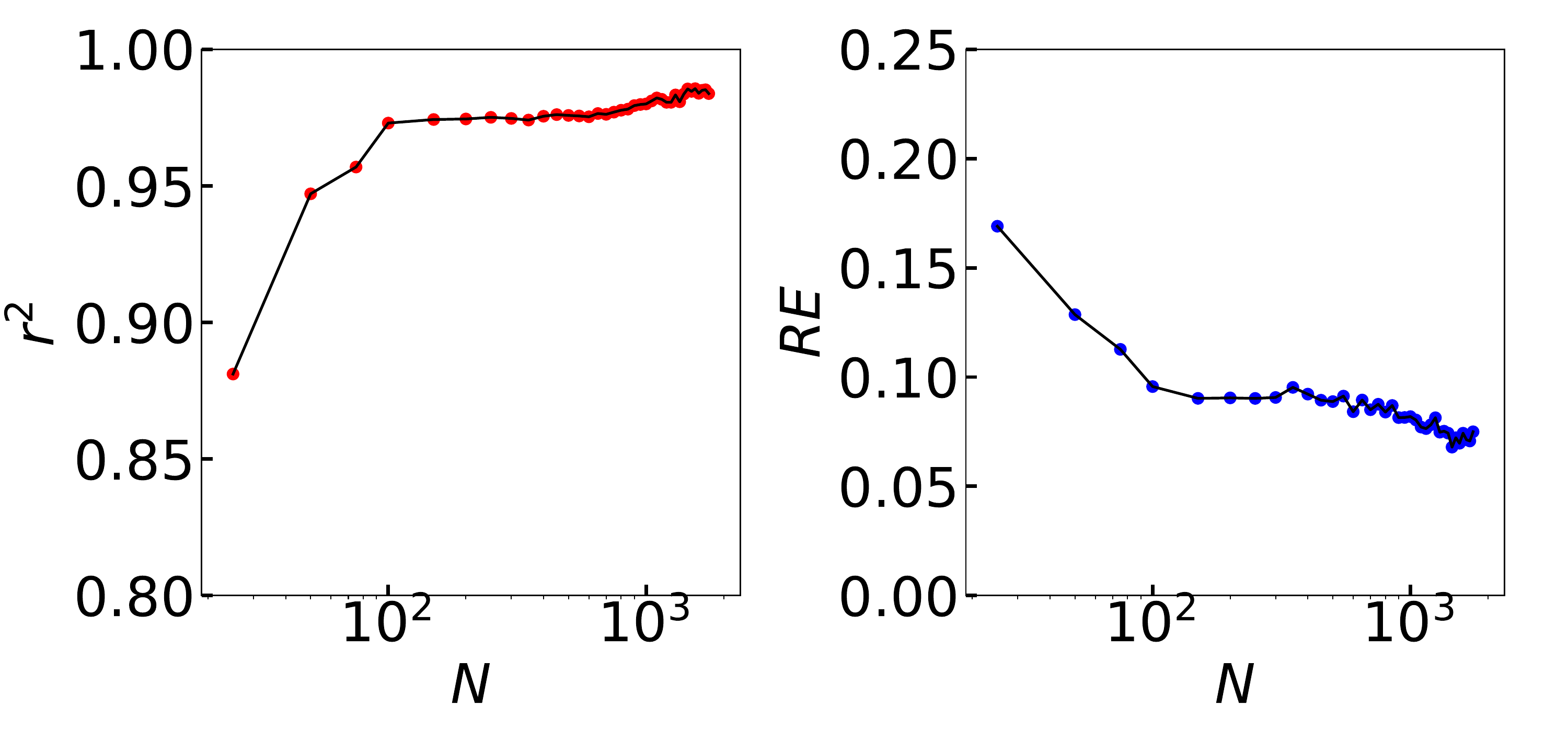}
\caption{Learning curve for the NN based ML model: (left) Accuracy ($r^2$) and (right) relative error ($RE$) of the test data set as a function of number of data ($N$) in the training set where the dimensionality of the output vector $n_0=40$.}
\label{fig:s1}
\end{figure}

\section{Random Forest Learning Curve}
\setcounter{figure}{0}
\renewcommand{\thefigure}{B.\arabic{figure}}

\begin{figure}[h!]
\centering
\includegraphics[width=0.9 \columnwidth]{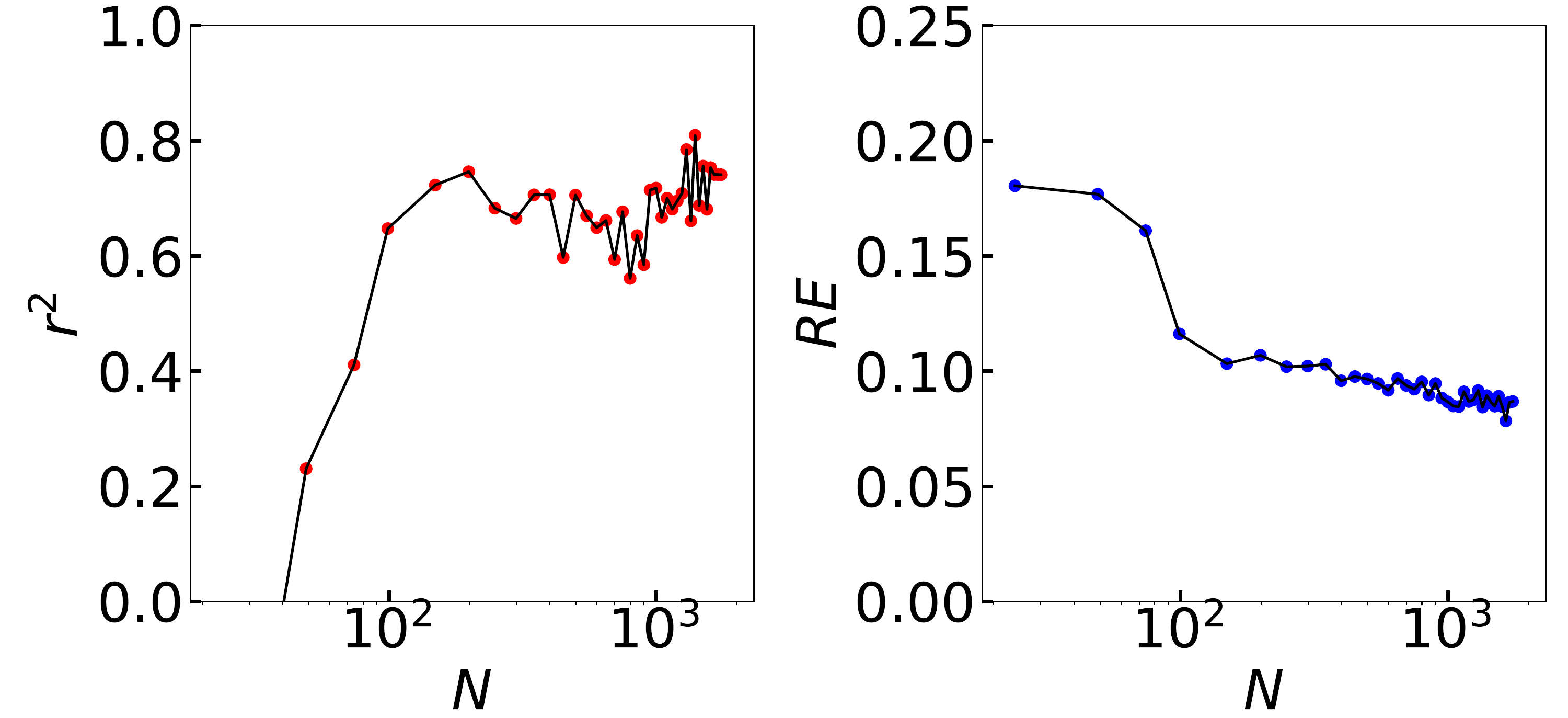}
\caption{Learning curve for the RF  model : (left) Accuracy ($r^2$) and (right) relative error ($RE$) of the test data set as a function of number of data ($N$) in the training set.}
\label{fig:s2}
\end{figure}

To check the generality of the machine learning approach, we also explored {\textit{Random Forest (RF)}} ML model on our data set and generated the following (see Fig. \ref{fig:s2}) learning curve. A quick comparison (see Fig.~\ref{fig:s1} and Fig.~\ref{fig:accuracy} in main text) between different learning curves revel that {\textit{Random Forest (RF)}} model gives rather relatively poor ML model in comparison to the neural network based models. We used Scikit-learn\cite{scikit-learn} python module for the implementation of the {\textit{Random Forest (RF)}} algorithm.   
\\

\section{NN Training  Curve}
\setcounter{figure}{0}
\renewcommand{\thefigure}{C.\arabic{figure}}
The training curves (both accuracy $r^2$ and relative error $RE$) for our neural network based machine learning model are presented below for two different cases : $1800$ training data points (case A; see Fig.~\ref{fig:s3}) and $100$ training data points (case B; see Fig.~\ref{fig:s4}). In both the cases, 100 test data points were used. In case A (1800 training data points), optimal fitting of the test data set is achieved for $\sim 10^{3}$ iteration (epoch) steps while  only $\sim 10^{2}$ epoch is required to achieve optimal fitting for case B (100 training data points). Above these optimal epoch value, the training data set is overfitted reducing the fitting accuracy of the test data as shown in both Fig. \ref{fig:s3} and \ref{fig:s4}. 

\begin{figure}[h!]
\centering
\includegraphics[width=0.9 \columnwidth]{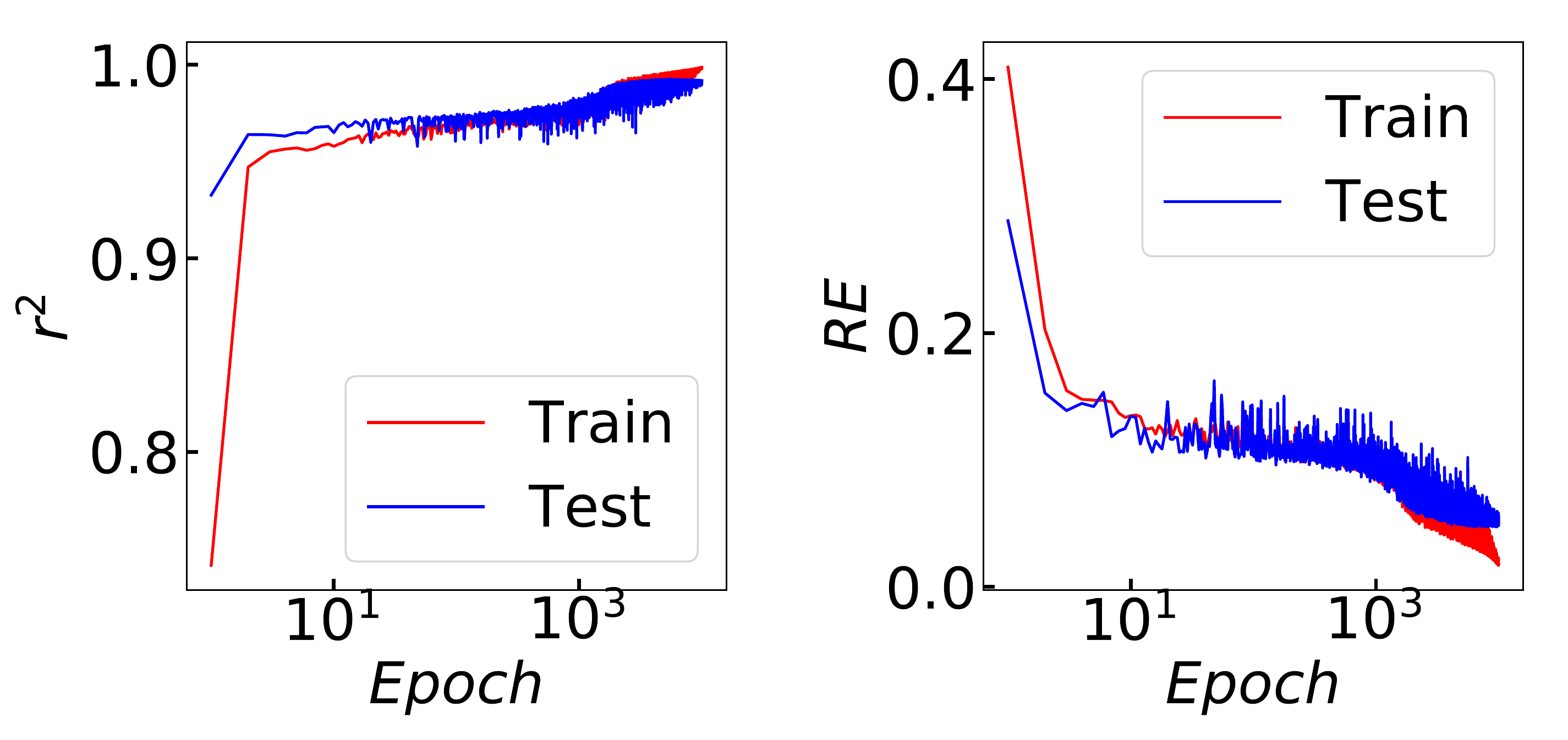}
\caption{(left) Accuracy ($r^2$) and (right) relative error (RE) of the test data set, as a function of epoch for Case A (see text).}
\label{fig:s3}
\end{figure}

\begin{figure}[h!]
\centering
\includegraphics[width=0.9 \columnwidth]{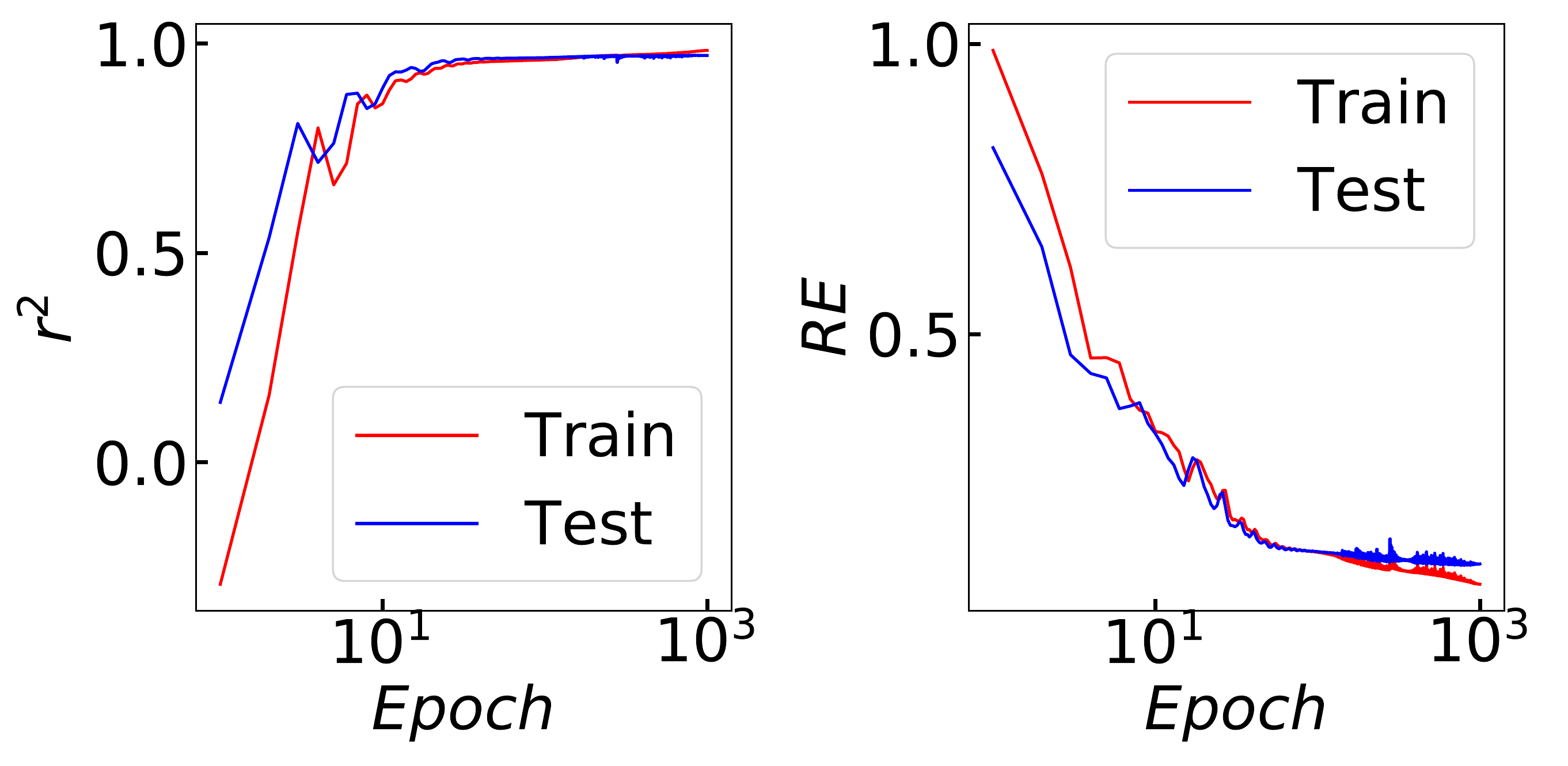}
\caption{(left) Accuracy ($r^2$) and (right) relative error (RE) of the test data set, as a function of epoch for Case B (see text).}
\label{fig:s4}
\end{figure}

\bibliography{ml}

\end{document}